\documentclass[prl,reprint,amssymb]{revtex4-1}
\usepackage{amssymb,amsfonts,amsmath}
\usepackage{array}
\usepackage{dcolumn}
\usepackage{longtable}
\usepackage{hyperref}
\usepackage{float}
\usepackage{gensymb}
\usepackage{bbm}
\usepackage{bm}
\usepackage{graphicx}
\usepackage{datetime}
\usepackage{pifont}
\usepackage{comment}
\usepackage{dsfont}
\usepackage{color}

\def\d{\text{d}}

\begin{document}
\title{Contractile units in disordered actomyosin bundles arise from F-actin buckling}

\author{Martin Lenz$^1$, Todd Thoresen$^2$, Margaret L. Gardel$^{1,2,3}$, and Aaron R. Dinner$^{1,2,4,}$}
\email{dinner@uchicago.edu}
\affiliation{$^1$James Franck Institute, $^2$Institute for Biophysical Dynamics, $^3$Department of Physics, $^4$Department of Chemistry, University of Chicago, Chicago IL 60637, USA}

%%%%%%%%%%%%%%%%%%%%%%%%%%%%%%%%%%%%%%%%%%%%%%
\begin{abstract}
Bundles of filaments and motors are central to contractility in cells. The classic example is striated muscle, where actomyosin contractility is mediated by highly organized sarcomeres which act as fundamental contractile units.  However, many contractile bundles \emph{in vivo} and \emph{in vitro} lack sarcomeric organization.
Here we propose a model for how contractility can arise in actomyosin bundles without sarcomeric organization and validate its predictions with experiments on a reconstituted system.  In the model, internal stresses in frustrated arrangements of motors with diverse velocities cause filaments to buckle, leading to overall shortening.  We describe the onset of buckling in the presence of stochastic actin-myosin detachment and predict that buckling-induced contraction occurs in an intermediate range of motor densities. We then calculate the size of the ``contractile units'' associated with this process.
Consistent with these results, our reconstituted actomyosin bundles contract at relatively high motor density, and we observe buckling at the predicted length scale.
\end{abstract}

\maketitle

\begin{figure}[t]
\begin{center}
\includegraphics[width=8.7cm]{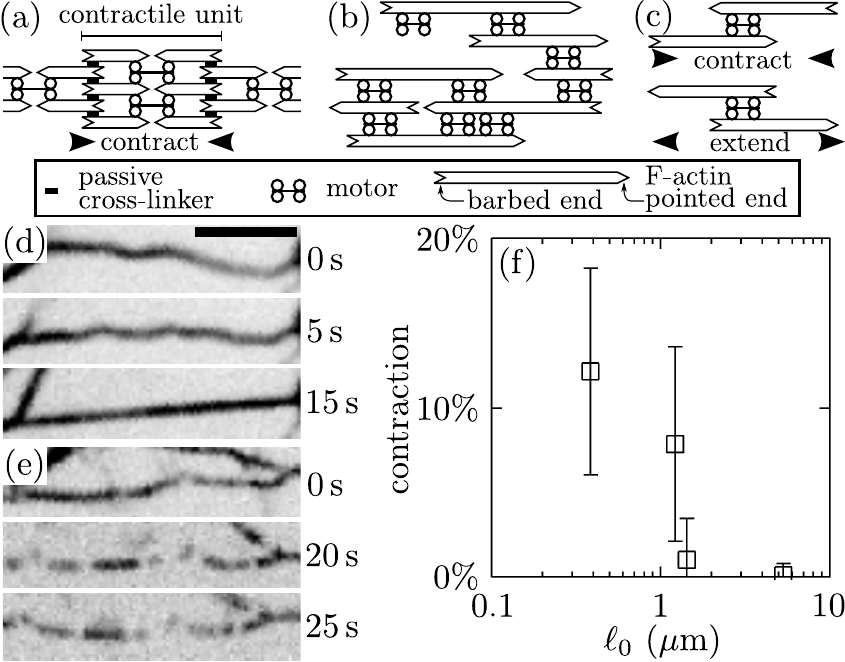}
\caption{\label{fig:contraction}Contraction in actomyosin bundles. 
(a)~Sarcomeric structure as in striated muscle, where passive cross-linkers are attached near F-actin barbed ends. Motors move towards F-actin barbed ends, causing each contractile unit (sarcomere) to contract.
(b)~Bundle devoid of sarcomeric organization or passive cross-linkers, as in our experiments.
(c)~Motors and polar filaments induce local contraction or extension depending on the geometry of their assembly. Overall contractility requires breaking the balance between these two effects.
(d)~Time-lapse images of a bundle comprised of F-actin and fluorescent myosin thick filaments (inverted contrast) with $\ell_0=540\,$nm. The initially wavy bundle becomes taut following the addition of $1\,\textrm{mM}$ ATP at $t=0\,\textrm{s}$, indicating contraction. Scale bar, $5\,\mu\textrm{m}$.
(e)~Similar experiment with $\ell_0=1.5\,\mu$m, showing no contraction. Scale bar as in (a). See also Movie~S1.
(f)~Bundle contraction as a function of $\ell_0$. Bars indicate standard deviation ($n\geqslant 25$).
}% Note: points in (f) have x-coordinates $\ell_0=0.386$, $1.22$, $1.43$ and $5.32\,\mu$m, respectively.
\end{center}
\end{figure}

Contractility arising from interactions between myosin molecular motors and actin filaments (F-actin) is used ubiquitously by cells to build tension and drive morphological changes \cite{Stricker:2010}.
Such force transmission from molecular to cellular length scales is well understood in striated muscle, where it critically relies on highly organized structures known as sarcomeres [Fig.~\ref{fig:contraction}(a)] \cite{Alberts:1998aa}.
However, many contractile actomyosin bundles found \emph{in vivo}, such as smooth muscle fibers \cite{Fay:1983}, graded polarity bundles \cite{Cramer:1997} and the contractile ring \cite{Carvalho:2009}, lack a sarcomeric organization. Most recently, we have shown that \emph{in vitro} bundles lacking apparent sarcomeric organization can also contract \cite{Thoresen:2011} [\emph{e.g.}, Fig.~\ref{fig:contraction}(b)]. In these disparate systems, contraction occurs with a well-defined contraction velocity per unit length, suggesting that contractile bundles can be meaningfully divided into elementary units that are arranged in series \cite{Carvalho:2009,Thoresen:2011,BementHerrera:19912005}. The mechanisms giving rise to such units in the absence of sarcomeric organization are not understood.

Much theoretical work on non-sarcomeric actomyosin assemblies posits contractility as a fundamental assumption, and predicts larger-scale effects such as polarity organization \cite{KruseYoshinaga:2003ab2010}, the appearance of topological defects \cite{KruseKruse:2004aa2005aa}, active stiffening \cite{MacKintosh:2008aa}, and oscillatory behavior in cells \cite{Salbreux:2007}. Models that address the microscopic origin of contractility assume that myosin motors dwell at the barbed ends of F-actin, thus acting as transient static cross-linkers \cite{MeanField}. This generates sufficient sarcomere-like organization to elicit contraction \cite{Zemel:2009}. Experimental evidence for this behavior is unfortunately lacking \cite{Supplement}, and it is thus important to investigate alternative mechanisms.

Here we demonstrate another route to contractility in non-sarcomeric bundles through theory and experiments. We first show experimentally that contraction in reconstituted actomyosin bundles is accompanied by F-actin buckling. Such buckling could provide a symmetry-breaking mechanism necessary for contraction \cite{Lenz:2011} [Fig.~\ref{fig:contraction}(c)], and we investigate the general consequences of asymmetric filament response theoretically by considering the build-up of forces in a bundle with randomly arranged motors. We predict that buckling, and thus contraction, can only be achieved in an intermediate range of motor density, due to the stabilizing effects of strong cross-linking at high density and of stochastic motor detachment at low density. This picture yields a characteristic length scale between two buckles, which provides a natural size for a contractile unit. These predictions are consistent with experimental observations, suggesting that buckling is necessary for contractility in non-sarcomeric actomyosin bundles.

To form reconstituted actomyosin bundles, we incubate F-actin with length $\ell_f\simeq 5\,\mu$m with smooth muscle myosin thick filaments of length $\simeq 300\,$nm in buffer lacking ATP such that thick filaments cross-link F-actin with high affinity. 
{While flexible motors have been considered as a basis for contraction} \cite{Liverpool:2009}, this is unlikely to apply here as thick filaments are significantly more rigid than F-actin.
The bundle lengths range from $10$ to $100\,\mu$m with $4$-$6$ F-actin per bundle cross-section, and no sarcomeric organization is observed \cite{Thoresen:2011}.  By varying the concentration of myosin filaments, the average spacing $\ell_0$ between two consecutive myosin filaments can be varied from $390\,$nm to $5.3\,\mu$m \cite{Thoresen:2011}. Perfusing buffer containing 1\,mM ATP causes bundles formed with high myosin density ($\ell_0=540\,\textrm{nm}$) to shorten by $\simeq 10\%$ rapidly ($100\textrm{-}600\,\textrm{nm}\cdot\textrm{s}^{-1}$) [Fig.~\ref{fig:contraction}(d) and Ref.~\cite{Supplement}; Movie~S1].  In contrast, contraction does not occur at low myosin density ($\ell_0=1.5\,\mu$m) [Fig.~\ref{fig:contraction}(e) and Movie~S1]. A sharp transition between those two behaviors is observed at $\ell_0=1.3\,\mu$m [Fig.~\ref{fig:contraction}(f)].

In considering these observations, it is important to recognize that actomyosin interactions can \emph{a priori} elicit extension just as well as contraction. As shown in Fig.~\ref{fig:contraction}(c), elementary bundles comprised of two F-actin (``filaments'') and one myosin thick filament (``motor'') containing numerous myosin heads contract when the motor is located in the vicinity of the pointed ends, but extend when it is close to the barbed ends. In sarcomeres, myosin is restricted towards the pointed ends of actin filaments, which favors contraction over extension [Fig.~\ref{fig:contraction}(a)].
By contrast, we show in Refs.~\cite{Supplement,Lenz:2011} that these two tendencies compensate and prevent overall contraction in non-sarcomeric bundles unless two specific conditions are fulfilled. First, the unloaded motor velocities need to have a certain dispersion among the motor population. Otherwise, motors merely induce filament translation without inducing overall bundle contraction \cite{Zemel:2009}. Such dispersion has been observed experimentally \cite{Yamada:1990}, and in our system likely arises from the variation of number of myosin heads in the thick filaments. In its presence, stresses build in the bundle as motors translate the filaments with different preferred velocities. While these stresses present a potential for bundle deformation, the disordered nature of the bundle implies that they have equal chances of being compressive or extensile. Thus the second condition is an asymmetric response of the filaments to such stresses, which breaks the symmetry between contraction and extension. Contraction occurs when filaments yield under compression while resisting extension \cite{Supplement,Lenz:2011}.

\begin{figure}[t]
\begin{center}
\includegraphics[width=8.7cm]{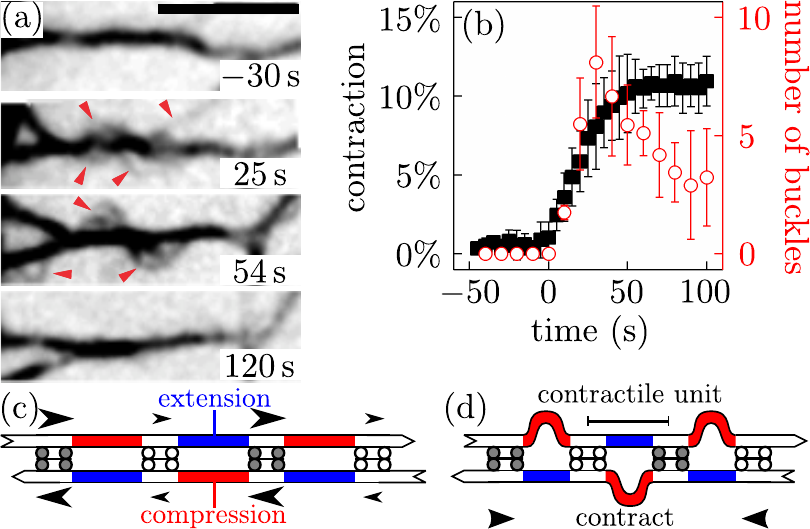}
\caption{\label{fig:buckles}Buckling in non-sarcomeric contractile actomyosin bundles. 
(a)~Time-lapse images of fluorescent actin (inverted contrast) showing F-actin buckling (\emph{arrowheads}) following the addition of $1\,\textrm{mM}$ ATP at $t=0\,\textrm{s}$. Scale bar, $5\,\mu\textrm{m}$. See also Movie~S2.
(b)~Relative contraction (\emph{filled squares}) and number of F-actin buckles (\emph{open circles}) as a function of time. Data shows mean $\pm$ sd averaged over $n=3$ bundles with $\ell_0\simeq 1\,\mu$m. 
(c)~The presence of fast (\emph{grey}) and slow (\emph{white}) motors generically induce compressive (\emph{red}) and extensile (\emph{blue}) stresses in filaments.
(d)~Buckling of the compressed filaments leads to an overall shortening of the bundle.
}
\end{center}
\end{figure}

{We next look for evidence of this behavior in our experiments. In this respect, we observe F-actin buckling coinciding with contraction [Fig.~}\ref{fig:buckles}{(a); Movie~S2]. This constitutes an extreme form of asymmetric response of the filaments, and thus enables contractility. Prior to ATP addition, compact bundles with aligned F-actin are observed.}
Upon ATP addition, the frequency of buckles increases rapidly during contraction, and then diminishes once contraction stops [Fig.~\ref{fig:buckles}(b)]. These F-actin buckles are dynamic, with their amplitude, curvature and location changing over time.

Qualitatively, the relationship between buckling and contraction can be understood as follows. Consider two antiparallel filaments interacting through several different motors with distinct speeds [Fig.~\ref{fig:buckles}(c)]. As motors start to move relative to the filaments, stresses build in sections of the filament flanked by motors with different speeds. When the flanking motor proximal to the barbed end is faster than that proximal to the pointed end, compression arises. When it is slower, tension arises. Following buckling of the compressed filament sections, fast motors are free to move quickly while the others move slowly. This results in the growth of the compressed sections and shrinkage of the extended ones, and thus in overall bundle contraction [Fig.~\ref{fig:buckles}(d)]. 
The region centered around each buckle thus plays the role of a contractile unit, whose typical size is equal to the distance $\ell_B$ between two buckles.

In this picture, the contractile behavior of the bundle hinges on the ability of the motors to induce filament buckling. At high motor density, we expect the bundle to be so strongly cross-linked that buckling becomes impossible despite the sizable stresses induced by a large number of motors. 
At low motor density, we expect 
that stochastic detachment of the motors undermines stress build-up and thus prevents buckling. Here we present a mathematical model to predict the range of myosin densities enabling contraction and the contractile unit length $\ell_B$. These results are then compared with the observations in Figs.~\ref{fig:contraction} and \ref{fig:buckles} to validate the proposed contraction mechanism.

\begin{figure}[t]
\begin{center}
\includegraphics[width=8.7cm]{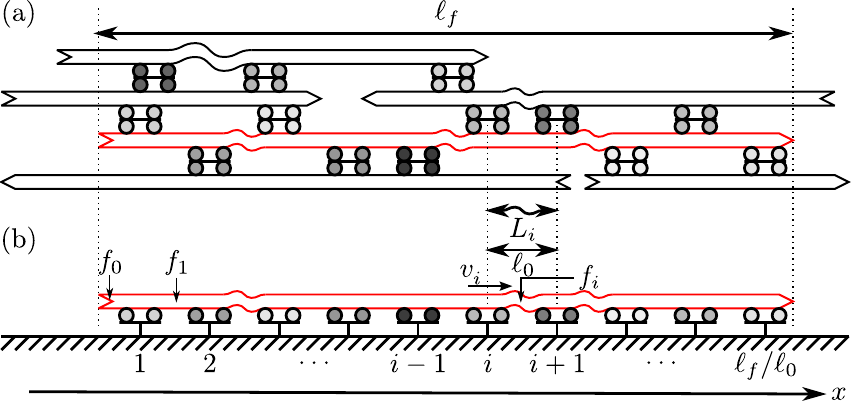}
\caption{\label{fig:model}
Stress build-up in bundles with non-identical motors.
(a)~In a bundle with motors having non-identical velocities (\emph{shades of grey}), filaments of lengths $\approx\ell_f$ are subjected to random motor forces at points $\approx\ell_0$ apart distributed throughout their length.
(b)~Prior to buckling, the environment of a filament of interest (\emph{red}) can be approximated by a collection of immobile motors (\emph{shades of grey}) \cite{Supplement}.
}
\end{center}
\end{figure}

The key assumptions of our model are that (1)~motors
have a dispersion in their unloaded velocities,
(2)~a section of filament between two motors buckles above a certain threshold force $F_B$, and (3)~motors
{intermittently}
detach from the filaments, thus
allowing
local stress relaxation. We consider a bundle of weakly deformed filaments and ask whether the forces developing within it are sufficient to induce buckling [Fig.~\ref{fig:model}(a)].

To this end,
we focus on a single filament of length $\ell_f$ and approximate its surroundings by an effective medium composed of immobile \cite{Supplement} point-like motors separated by a distance $\ell_0\ll\ell_f$ [Fig.~\ref{fig:model}(b)]. This divides the filament into discrete sections, which we label by $i=0,\ldots,\ell_f/\ell_0$. We take into account the possibility that the filaments are not straight, but bend away from the $x$-axis, implying that the contour length $L_i$ of filament section $i$ can be larger than $\ell_0$. Defining $f_i$ as the tension of filament section $i$ ($f_i<0$ for a compressed filament section), we expand its force-extension relationship for small deformations:
\begin{equation}\label{eq:fil}
L_i=L_i(f_i=0)-cf_i,
\end{equation}
where $c>0$ is the filament compliance. We refer to the motor flanked by filament sections $i-1$ and $i$ as ``motor $i$'', and describe its operation by the simplified force-velocity relationship
\begin{equation}\label{eq:forcevel}
f_{i-1}-f_i=F_i-\chi v_i.
\end{equation}
Here $v_i$ denotes the local velocity of the filament at the location of motor $i$ 
and $\chi>0$ is the motor susceptibility. Eqs.~(\ref{eq:fil}) and (\ref{eq:forcevel}) yield a local relaxation time scale $\tau_r=\chi c/2$. The time-independent stall force of motor $i$ is denoted by $F_i$ in Eq.~(\ref{eq:forcevel}), and is {drawn}
from a random distribution satisfying
\begin{equation}\label{eq:stall}
\overline{F_i}=F_S
\quad\textrm{and}\quad
\overline{F_iF_j}-\overline{F_i}\,\overline{F_j}=\delta F_S^2\delta_{ij},
\end{equation}
where bars denote averages over the motor distribution. 
As a result, different motors have different unloaded velocities $F_i/\chi$ as required for contraction. Owing to
the conservation of filament mass:
\begin{equation}\label{eq:conservation}
\frac{\d L_i}{\d t}=v_i-v_{i+1}.
\end{equation}
Finally, a motor bound to several filaments as in Fig.~\ref{fig:model}(a) can transiently detach from one while still holding onto the others \cite{Supplement}. We thus let each motor $i$ randomly detach from the filament with a constant rate $1/\tau_d$. Following detachment, local filament stresses relax instantaneously, yielding $f_i=f_{i-1}=(f_{i}+f_{i-1})/2$. The motor then reattaches after a time much shorter than $\tau_r$ and $\tau_d$ \cite{Supplement}. We denote by $\langle\ldots\rangle$ the average over the Poisson process of motor detachment.

We obtain the space and time evolution of the filament tension $f(x,t)$ in the continuum limit $i\rightarrow x/\ell_0$ by combining Eqs.~(\ref{eq:fil}-\ref{eq:conservation}) and averaging over motor detachment \cite{Supplement}:
\begin{equation}\label{eq:evolution}
\partial_t\langle f\rangle-D\partial_x^2\langle f\rangle=({\ell_0}/{2\tau_r})\partial_x F,
\end{equation}
where $D={\ell_0^2}({\tau_r^{-1}}+{\tau_d^{-1}})/2$. The right-hand-side of Eq.~(\ref{eq:evolution}) involves the spatial gradient of the stall force $F(x)$, reflecting the fact that non-identical motors lead to force build-up. This effect competes with the relaxation of filament forces through motor detachment, which 
enters through
the diffusion term $D\partial_x^2\langle f\rangle$. 

\begin{figure}[t]
\begin{center}
\includegraphics[width=8.7cm]{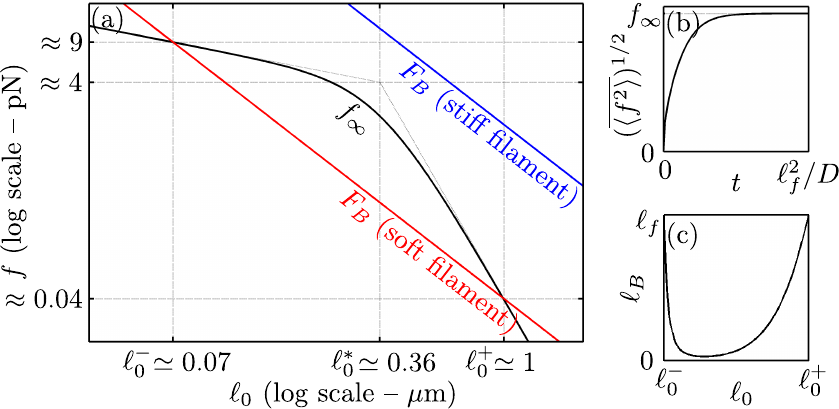}
\caption{\label{fig:regimediagram}
Model predictions for filament force build-up. 
(a)~\emph{Black line}: Steady-state filament force $f_\infty$ as a function of motor spacing $\ell_0$ as in Eq.~(\ref{eq:longtime}). For $\ell_0\ll\ell_0^*$ and $\ell_0\gg\ell_0^*$, $f_\infty\propto\ell_0^{-1/2}$ and $\ell_0^{-9/2}$, respectively. \emph{Colored lines}: buckling force $F_B\propto\ell_0^{-2}$.
(b)~Typical filament force $(\overline{\langle f^2\rangle})^{1/2}$ as a function of time \cite{Supplement}.
(c)~Contractile unit size $\ell_B$ as a function of $\ell_0$ as in Eq.~(\ref{ellB}) ($\ell_f\simeq 5\,\mu$m).
}
\end{center}
\end{figure}

An initially relaxed filament [$f(x,t=0)=0$] experiences a vanishing average force $\overline{\langle f\rangle}(x,t)=0$ throughout its dynamics. To quantify the magnitude of the motor-induced stress, we thus calculate the rms filament force $(\overline{\langle f^2\rangle})^{1/2}$ \cite{Supplement} and find that it increases monotonically from zero at $t=0$ to
\begin{equation}\label{eq:longtime}
f_\infty= \frac{1}{2\sqrt{3}}\left(\frac{\ell_f}{\ell_0}\right)^{1/2}\frac{\delta F_S}{1+\tau_r/\tau_d}.
\end{equation}
at $t=\infty$ [Fig.~\ref{fig:regimediagram}(a-b)]. We next estimate the dependence of the ratio $\tau_r/\tau_d$ on the experimentally accessible parameter $\ell_0$.
A Worm-Like Chain model for filament elasticity yields $c\approx {\ell_0^4}/{k_BT\ell_p^2}$, where $\ell_p$ is the filament persistence length \cite{Odijk:1995}, and we approximate $\chi\approx F_S/v$, where $v$ is a characteristic motor velocity. This implies $\tau_r/\tau_d\approx ({\ell_0}/{\ell_0^*})^4$, with $\ell_0^*=({k_BT\ell_p^2v\tau_d}/{F_S})^{1/4}$. We can thus distinguish two regimes for the steady-state force $f_\infty$ [Fig.~\ref{fig:regimediagram}(a)]. For $\ell_0\ll\ell_0^*$, detachment events are rare compared to the time $\tau_r$ needed for the force to recover from such an event, and $f_\infty$ is not affected by them. For $\ell_0\gg\ell_0^*$, $f_\infty$ quickly decreases with increasing $\ell_0$ as detachment becomes much faster than recovery.

Up to a prefactor of order one, contraction proceeds as in Fig.~\ref{fig:buckles}(c-d) if $f_\infty>F_B\approx {k_BT\ell_p}/{\ell_0^2}$ \cite{Odijk:1995}. Comparing $f_\infty$ to $F_B$ as in Fig.~\ref{fig:regimediagram}(a), we find a threshold stiffness above which buckling cannot occur (as exampled by the \emph{blue line}). Reasonable values for our actomyosin system are $\ell_p\simeq10\,\mu$m, $v\simeq 200\,\textrm{nm}\cdot\textrm{s}^{-1}$, $\delta F_S\approx F_S\simeq 1\,$pN and $\tau_d\simeq 200\,$ms based on the typical time scales involved in the myosin mechanochemical cycle \cite{Supplement}. These values put us in the soft filament regime defined by $\ell_p\ll \delta F_S^4 L_f^2 (v\tau_d)^{3/2}/k_BT^{5/2}F_S^{3/2}\simeq 20\,\textrm{cm}$ (\emph{red line}). In this regime, the lines representing $F_B$ and $f_\infty$ intersect at
\begin{subequations}\label{eq:ell0s}
\begin{eqnarray}
\ell_0^-&=&({k_BT\ell_p}/{\delta F_S\ell_f^{1/2}})^{2/3} \simeq 70\,\textrm{nm},\\
\ell_0^+&=&({\ell_f^{1/2}v\tau_d\ell_p\delta F_S}/{F_S})^{2/5} \simeq 1\,\mu\textrm{m},
\end{eqnarray}
\end{subequations}
meaning that buckling and contraction occur for
$\ell_0^-<\ell_0<\ell_0^+$.
This range reflects the fact that strong cross-linking ($\ell_0<\ell_0^-$) suppresses buckling while sparse motors ($\ell_0>\ell_0^+$) are undermined by stochastic detachment.
While the regime $\ell_0\simeq\ell_0^-$ is not accessible experimentally, the predicted value for $\ell_0^+$ is strikingly similar to the motor spacing at which the breakdown of contraction is observed in Fig.~\ref{fig:contraction}(f) ($1.3\,\mu$m), suggesting that the proposed mechanism is a good description of our experiments.  

To characterize the contractile units resulting from this mechanism when $\ell_0^-<\ell_0<\ell_0^+$, we turn to the transient regime leading up to filament buckling.
The filament force profile as a function of $x$ is initially flat, and subsequently coarsens into a random walk for $t=+\infty$. According to Eq.~(\ref{eq:evolution}), this coarsening occurs diffusively with diffusion coefficient $D$. The typical filament forces at time $t\ll \ell_f^2/D$ are thus of order $f_\infty(\sqrt{Dt}/\ell_f)^{1/2}$. We denote the time that this force reaches the buckling threshold $F_B$ by $t_B$, following which contraction proceeds as in Fig.~\ref{fig:buckles}(c-d) and the coarsening dynamics is interrupted. The distance between buckles at $t_B$ thus yields the contractile unit size
\begin{equation}\label{ellB}
\ell_B\approx \sqrt{Dt_B} \approx\frac{\ell_p^2}{\ell_0^3}\left(\frac{k_BT}{\delta F_S}\right)^2\left(1+\frac{\tau_r}{\tau_d}\right)^2.%\\
\end{equation}
As illustrated in Fig.~\ref{fig:regimediagram}(c), $\ell_B$ is typically in the micrometer range, in agreement with the observations of Fig.~\ref{fig:buckles}(a) and the findings of Ref.~\cite{Thoresen:2011}.

Because of compensating effects between contractile and extensile motor-filament configurations, the familiar framework involving rigid filaments and identical motors commonly used to describe striated muscle contraction is not suited to study actomyosin bundles lacking sarcomeric organization. Here, we put forward an alternative mechanism based on our observation of buckling.  The buckling arises from the nonlinear elastic response of F-actin \cite{Berro:2007} and dispersion in the speeds of myosin motors \cite{Tanaka:1992}.  F-actin buckling has previously been invoked to explain contraction qualitatively \cite{Silva:2011}. Addition of passive cross-linkers, which are formally equivalent to immobile motors, would reinforce a dispersion of motor velocities and promote contraction.

The order-of-magnitude agreement between theory and experiments with respect to the size of contractile units and the critical myosin concentration required for contraction suggests that our current analysis offers a good description of the onset of bundle contractility. 
Our conclusions are robust to inclusion of features such as inhomogeneous motor spacings $\ell_0$ and force dependence of the motor detachment rate \cite{Supplement}. {Our mechanism is a general one and applies to any one-dimensional system of polar filaments and motors. Further experiments and theory are needed to better understand the molecular basis for motor inhomogeneities and filament asymmetric response in the myriad of non-sarcomeric organizations found \emph{in vivo}. }

\begin{acknowledgments}
We thank Yitzhak Rabin and Tom Witten for useful discussions.  This work was supported by NSF DMR-MRSEC 0820054, NIH P50 GM081892 and NIH DP10D00354.
\end{acknowledgments}

\end{document}